\documentclass[useAMS]{mn2e}
\usepackage{times}

\def\eq#1 {eq.~(\ref{eq:#1})}

\def\etal{{\it et al.\ }}

\def\eg{{\it e.g.},}
\def\ie{{\it i.e.},}

\def\ltsima{$\; \buildrel < \over \sim \;$}
\def\lsim{\lower.5ex\hbox{\ltsima}}
\def\gtsima{$\; \buildrel > \over \sim \;$}
\def\gsim{\lower.5ex\hbox{\gtsima}}
\def\ga{\mathrel{\hbox{\rlap{\hbox{\lower4pt\hbox{$\sim$}}}\hbox{$>$}}}}
\def\la{\mathrel{\hbox{\rlap{\hbox{\lower4pt\hbox{$\sim$}}}\hbox{$<$}}}}

\def\kms{\,{\rm km\,s{^{-1}}}}
\def\hmpc{\,h{^{-1}}{\rm Mpc}}

\def\la{\dangler}

\def\pmb#1{\setbox0=\hbox{#1}%
 \kern-.025em\copy0\kern-\wd0
 \kern.05em\copy0\kern-\wd0
 \kern-.025em\raise.0433em\box0}

\def\br{{\bf r}}
\def\bv{{\bf v}}

\def\hmpc{\, h^{-1} {\rm Mpc}}

\def\3hmpc{\, ( h^{-1} {\rm Mpc})^3}
\def\kms{\, {\rm km\,s^{-1}}}

\def\bk {{ \bf k}}

\def\bx {{ \bf x}}

\def\r0p { r{_0^\prime}}

\title[The Gaussian cell two-point ``energy-like'' equation] 
{The Gaussian cell two-point ``energy-like'' equation: Application to
  large scale galaxy redshift and peculiar motion surveys}

\author[Zaroubi \& Branchini]{Saleem Zaroubi$^1$ and Enzo
  Branchini$^2$\\$^1$Kapteyn Astronomical Institute, University of
  Groningen, Landleven 12, 9747 AG Groningen, The Netherlands\\$^2$
  Dipartimento di Fisica dell$`$ Universit\`{a} degli Studi $``$Roma
  TRE$''$, Via della Vasca Navale 84, I-00146, Roma, Italy}

\begin{document}

\maketitle
\begin{abstract}
We introduce a simple linear equation relating the line-of-sight
peculiar velocity and density contrast correlation functions. The
relation, which we call the {\it Gaussian cell two-point
``energy-like'' equation}, is valid in the distant-observer-limit and
requires Gaussian smoothed fields. In the variance case, \ie\ at zero
lag, the equation is similar in its mathematical form to the
Layzer-Irvine cosmic energy equation.  $\beta$ estimation with this
equation from the PSC$z$ redshift galaxy survey and the SEcat
catalogue of peculiar velocities is carried out, returning a value of
$\beta = 0.44\pm0.08$. The applicability of the method for the 6dF
galaxy redshift and peculiar motions survey is demonstrated with mock
data where it is shown that beta could be determined with $\approx
5\%$ accuracy. The prospects for constraining the dark energy
equation of state with this method from the kinematic and thermal
Sunyaev-Zel'dovich cluster surveys are discussed. The equation is also
used to construct a nonparametric mass density power spectrum
estimator from peculiar velocity data.
\end{abstract}


\section{Introduction}
\label{introduction}

In the linear regime of the gravitational instability scenario the
underlying mass distribution is directly traced by the galaxy peculiar
velocities. Measurement of the radial component of the galaxy peculiar
velocities, the only component that one can easily observe, is carried
out with one of many available techniques, the most common among which
are the Tully-Fisher-like methods (\eg\ Tully \& Fisher 1977, Faber \&
Jackson 1976).  Normally, these methods exploit a well defined
intrinsic relation between two or more of the galaxy\footnote[1]{It
should be noted that some methods are not based on galaxy properties
and use other ``extra-galactic objects'', \eg\ supernovae-Ia.}
observed properties that facilitates establishing its actual distance
from the observer. The estimated distance is then used together with
the measured galaxy redshift, to determine the galaxy radial peculiar
velocity. Assuming an irrotational flow on large scales and the
knowledge of $\Omega_m$ (the cosmological mass density parameter), it
is straightforward to use the measured radial peculiar velocities to
recover the full underlying mass overdensity (Bertschinger \& Dekel
1989, Dekel, Bertschinger \& Faber 1990, Zaroubi 2002, Zaroubi \etal
2002).  In addition, the same mass-density could be probed by galaxy
redshift catalogues assuming a simple linear biasing scheme that
connects it to the spatial galaxy distribution (Kaiser 1984, Bardeen
\etal 1986). To date, galaxy redshift and peculiar motion surveys are
the main tools with which astronomers explore the distribution of
matter in the nearby universe.

Since the two types of data, galaxy peculiar velocity catalogues and
galaxy redshift surveys, probe the underlying mass distribution
comparing the two provides a simple and powerful test on the paradigm
of gravitational instability and gives a model independent measurement
of $\beta$, the ratio between the linear growth factor, $f(\Omega_m)$
($\approx \Omega_m^{0.6}$) and the linear biasing factor of the galaxy
population. In most cases the comparison is either performed by
deriving galaxy peculiar velocities from the galaxy density field and
confront them with the measured velocities point-by-point, an approach
usually called ``velocity-velocity'' comparison (e.g., Davis, Nusser
\& Willick 1996, Willick \& Strauss 1998, Zaroubi 2002, Zaroubi \etal
2002). Or by adopting the so called ``density-density'' approach in
which the velocity data is used to infer the full mass density field
(Bertschinger \& Dekel 1989, Zaroubi 2002) and compare it with the
galaxy distribution (\eg\ Sigad et al. 1998, Zaroubi \etal 2002). With
the exception of the POTENT algorithm (see \eg\ Sigad et al. 1998), all
the comparison methods yield a low value of $\beta$ consistent with
$\Omega_m\approx 0.3$ and bias factor of $\approx 1$.

Another approach to the comparison that doesn't fit into the two
general classes outlined earlier, is the one proposed by Juszkiewicz
\etal (1999) who start from the pair conservation equation (Peebles
1980) and evolve it further to the quasi-linear regime of
gravitational instability. In the pair conservation approach, which
yields a value of $\beta$ that is consistent with the one derived from
velocity-velocity analyses (Ferreira \etal 1999, Feldman \etal 2003),
a relation between the mean pairwise velocity at a certain separation
and the density correlation function is derived. The comparison in
this approach is significantly simplified by avoiding the spatial
point-by-point matching required in previous methods thus reducing the
noise involved. The Juszkiewicz \etal (1999) approach is similar to
the one developed in this paper except that we are interested in the
variance of the peculiar velocity at a given smoothing scale rather
than the pairwise peculiar velocity at a given separation.

In this study, we derive a very simple, model independent and linear
relation, valid in the distant observer limit, between the overdensity
and peculiar velocity two point correlation functions assuming Gaussian
smoothing. The method is first used to construct a nonparametric
estimator of the mass-density power spectrum. Then the paper
concentrates on the relation between the variance (2-point correlation
at zero distance) of the two fields. This relation basically reduces
the comparison between the catalogues to two numbers allowing a robust
extraction of the parameter $\beta$. The proposed equation is especially
suited to future peculiar velocity data sets like, the 6dF which will
measure the the peculiar velocities of $15,000$ galaxies with their
$D_n-\sigma$ relation up to $150$~Mpc/h distance.

Currently, the main sources of error in the redshift-peculiar motion
comparison are the large random and systematic uncertainties carried
by the peculiar motion measurements, for example the Tully-Fisher-like
relations has an inherent uncertainty of $\approx 15-20\%$ of the
distance. Data obtained with more accurate distance indicators do
exist (Tonry 1991, Riess, Press \& Kirshner 1995), unfortunately
however, either they are not at significant distances, \eg\ the
Surface-Brightness-fluctuations method, or they reach large distances
but are too sparse, \eg\ the Supernovae-Ia data.  In the future, by
using the thermal and kinematic Sunyaev-Zeldovich (SZ) effect (Sunyaev
\& Zeldovich 1972), both the accuracy of the peculiar velocity
measurement and the spatial coverage of the data are expected to
increase dramatically where the uncertainty is expected to amount to
an absolute error of $\approx 150 \kms$ (\eg Diaferio \etal 1994) and
the number of observed objects to reach $\approx 10^4$ clusters. The
main difficulty in this case, assuming a reasonable control over the
systematics, will be posed by the large mean separation between the
galaxy-clusters observed with the SZ effect.

The power of the method proposed here is that it reduces the
contribution of the measurement noise and sparseness of the sample to
a bare minimum.  The paper is organized as follows:
section~\ref{theory} presents the main theoretical formulae. In
section~\ref{SEcatPSCz} the method is applied to the PSC$z$ redshift
galaxy catalogue and the SEcat peculiar velocity survey. In
section~\ref{6dF} the applicability of the method to future surveys,
\eg\ the 6dF galaxy survey and the kinematic and thermal SZ cluster
survey, is discussed.


\section{Theoretical Derivations}
\label{theory}

In this section, we first derive the main theoretical relation
(subsection~\ref{2point}) and show how it could be used to estimate
the matter power spectrum from peculiar velocity data
(subsection~\ref{powerspec}). Then in subsection~\ref{estimator} the
variance component of the main relation is used in order to estimate
the value of $\beta$ from comparison between galaxy redshift surveys
data and peculiar velocity data. The $\beta$ measurement error for a
typical case is derived in subsection~\ref{noise}. The derivation is
performed within the framework of linear gravitational instability,
under the assumption of statistical homogeneity and isotropy.

\subsection{The basic relation}
\label{2point}

Consider a radial peculiar velocity field $v_{los}({\bf r})$ measured
in a very distant patch of the sky smoothed with a Gaussian window
function with scale $R_s$, $W_{R_s}(r)=(2\pi {R_s}^2)^{-3/2}
\exp(-{r^2 / 2 R^2_s})$. In the limit of $R_c \ll R$, where $R_c$ is
the correlation radius of peculiar velocities and $R$ is the distance
of the patch from the observer.  A smoothed radial field within a
given observed volume can be written as,
\begin{equation}
v_{los}^{S}(\bx) = {-\imath\beta { H_0}\over (2\pi)^3} \int
{\hat\br_{los}\cdot \bk \over k^2} \delta_k W_{R_s}(k) \exp(-i \bk
\cdot \bx) d^3k,
\label{eq:vzsmooth}
\end{equation}
where the superscript $S$ refers to values smoothed with a Gaussian
kernel of radius $R_s$, $\hat\br_{los}$ is a unit vector along the
line-of-sight, and $W_{R_s}(k)$ is Fourier transform of the smoothing
kernel.

The two-point correlation function of the Gaussian smoothed
line-of-sight galaxy peculiar velocity is:
\begin{eqnarray}
\langle \xi_{\mathbf{v}}^{los}(\br; R_s) \rangle & \equiv & \langle
v_{los}^{S}(\bx) v_{los}^{S}(\bx+\br) \rangle \\ & = & {\beta^2 {
H_0^2}\over (2\pi)^3} \int {(\hat\br_{los} \cdot \bk)^2\over k^4} P_k
W^2_k e^{-i \bk \cdot \br} d^3k.
\label{eq:v2pt}
\end{eqnarray}
Where $P_k$ is the mass density power spectrum and $\br$ is the radius
vector separating between any two points. Since there are two
independent directions that appear in eq.~\ref{eq:v2pt} one can't
invoke symmetry arguments in order to proceed. However, since in the
distant-observer-limit the line-of-sight direction is approximately
constant across the observed volume and independent of the direction
of $\br$, one can average over all possible directions of $\br$
relative to $\bk$ (arbitrary direction) by integrating
equation~\ref{eq:v2pt} with $\frac{1}{2} \int_{-1}^1 d\mu$, where
$\mu$ is cosine the angle between the two vectors $\br$ and $\bk$:
\begin{equation}
\langle \xi_{\mathbf{v}}^{los}(\br; R_s) \rangle_{\mu} = 
       {{(\beta H_0)}^2\over (2\pi)^3} \int{(\hat\br_{los} \cdot
      \bk)^2} {P_k\over k^4} W^2_k\ j_0(kr) d^3k. 
\end{equation}
Where $\langle \rangle_\mu$ is an average over statistical ensemble
and over $\mu$, $j_0(kr)$ is the zero order Spherical Bessel function
and $r=|\br|$

Assuming statistical isotropy for the velocity field, \ie\ symmetry
between the line of sight and the other two orthogonal directions one
obtains the following equation:
\begin{equation}
\langle \xi_{\mathbf{v}}^{los}(\br; R_s) \rangle_{\mu} = {\beta^2 { H_0^2}\over
      3(2\pi)^3} \int{P_k\over k^2}  W^2_k(R_s)j_0(kr) d^3k.
\end{equation}
With the factor $3$ coming from the symmetry argument.  Now to the
last step in the calculation, for a Gaussian smoothing kernel, \ie\
$W_{R_s}(k) = exp(-k^2R^2_s/2)$, the derivative of the line of sight
velocity two point correlation function with respect to $R_s$, yields:
\begin{eqnarray}
{d\langle \xi_{\mathbf{v}}^{los}(r; R_s) \rangle_\mu \over dR_s} & = & -{2 \over 3}
\beta^2
{ H_0^2} R_s \int P_k W^2_k j_0(kr)\frac{d^3k}{(2\pi)^3} \\ & = & -{2 \over 3} \beta^2 { H_0^2} R_s \xi(r; R_s).
\label{eq:v2ptd2pt}
\end{eqnarray}
Here $\xi(r; R_s)$ is the two point correlation function of the
smoothed densities. Notice that eq.~\ref{eq:v2ptd2pt} can only be
obtained when $d W_{R_s}(k)/dR_s \propto k^2 W_{R_s}(k)$, strictly
valid only with Gaussian smoothing kernel\footnote[2]{There are other
functions that satisfy this relation but they do not satisfy the
requirements of smoothing kernels}.

Obviously, for a given 3 dimensional peculiar velocity field the two
point correlation function of the Gaussian smoothed velocity, $\bv^S$,
is related to its density counterpart through the equation,
\begin{eqnarray}
{d\langle \xi_{\mathbf{v}}(r; R_s) \rangle \over dR_s} & \equiv & \langle
\bv^S(\bx)\cdot\bv^S(\bx+\br)\rangle \\
& = &   -2 \beta^2 { H_0^2} R_s \xi(r; R_s).
\label{eq:3dv2pt}
\end{eqnarray}
which is similar to equation~\ref{eq:v2ptd2pt} without the factor of
$3$ and with no need for averaging over $\mu$. 

It might be easier to interpret equation~\ref{eq:v2ptd2pt} in its
integral form, where the integral is performed over the smoothing
radius. Let $R_1$ and $R_2$ the two smoothing radii that bound our
integral, therefore,
\begin{equation}
\frac{1}{2}\langle \xi_{\mathbf{v}}^{los}(r; R_s) \rangle_\mu\mid_{R_1}^{R_2}
= -{\beta^2\over 3} { H_0^2} \int_{R_1}^{R_2}{\xi(r;
R_s)\over R_s} {d^3R_s \over 4\pi}.
\label{eq:energy}
\end{equation}
In the $r=0$ limit, the left-hand side of equation~\ref{eq:energy}
describes the mean change in the kinetic energy associated with the
smoothed velocity due to the variation of the smoothing radius,
whereas the right-hand side depicts the 3-dimensional integral of the
density variance of the smoothed field over the smoothing scale. The
right-hand side term is very similar to the normal potential energy
except that $R_s$ does not represent a proper distance between
points. In other words, the variation in the kinetic-like energy comes
from the ``potential-energy-like'' behavior of the modes corresponding
to the scales between $R_1$ and $R_2$, with exponentially decreasing
contributions from larger and smaller scale modes. If $r\ne 0$ then
the interpretation is not as simple but still the left- and right-hand
sides correspond to a sort of a {\it two-point kinetic-energy} and
{\it potential-energy}, respectively, with the same scales
contributing to the modification as before.  Therefore,
equation~\ref{eq:energy} is an ``energy equation'' of sorts as it
describes the {\it two point} ``potential-like'' and ``kinetic-like''
energy partition within a Gaussian window function. Dimensional
arguments significantly restrict the mathematical form of the relation
between the velocity and the density 2 points correlation functions.
Therefore, it is not surprising that the theoretical relation shown by
equation~\ref{eq:v2ptd2pt} is very similar to the Irvine-Layzer cosmic
energy equation which describes how the energy of the Universe is
partitioned between kinetic and potential energy (Irvine 1961, 1965;
Layzer 1963, 1966; see also Mo, Jing \& B\"orner 1997).

\subsection{The mass density power spectrum}
\label{powerspec}

The main approach currently used to measure the mass-density power
spectrum from peculiar velocity data is the likelihood method
introduced by Zaroubi \etal\ (1997; see also, Freudling \etal\ 1999
and Zaroubi \etal\ 2001) in which a theoretical power spectrum with
few free parameters and a noise model are assumed.  Since in this
method the data is forced to fit a specific power spectrum shape, an
inaccurate description of the noise model could propagate to large
scales and contaminate the measured power spectrum. Indeed the results
obtained with this method have been yielding unrealistically high
amplitude of the mass-density fluctuations power spectrum (consistent
with $\Omega_m > 0.6$). Therefore, direct nonparametric methods for
power-spectrum estimation from peculiar velocity data are needed.

The question we pose here is: can equation~\ref{eq:v2ptd2pt} be used
to directly estimate the mass power spectrum from peculiar velocity
data? In the ideal case in which the uncertainties in the measurement
are neglected and the data extends to infinity and samples the
universe very accurately, the answer is clearly yes.

A good point from which to start the derivation of the power spectrum
estimator is the relation between the power spectrum and the smoothed
density two point correlation function,
\begin{equation}
\xi(r;R_s) = {1 \over 2\pi^2} \int P_{k'} e^{-k^2R_s^2} j_0(k'r) k'^2 dk'.
\label{eq:corrps}
\end{equation}
Substituting this into equation~\ref{eq:v2ptd2pt} and integrating over
the variable $r$ from zero to $\infty$ after multiplying with $j_0(kr)
r^2$, one obtains the following relation,
\begin{equation}
{6 \pi \over H_0^2 R_s} \int_0^{\infty} {d\langle \xi_{\mathbf{v}}^{los}(r; R_s)
  \rangle_\mu \over dR_s} j_0(k r) r^2 dr = f^2 P_k e^{-k^2R_s^2}.
\label{eq:ps}
\end{equation}
Where the orthonormality of the spherical Bessel function is used
(\eg\ Arfken \& Weber 2002).  The left hand side of eq.~\ref{eq:ps} is
a quantity that can be directly measured from the velocity data;
whereas, the right-hand side shows the estimated quantity. Since we use
smoothed velocity data, the power spectrum is determined up to a
factor of $f^2(\Omega_m)$ and with a resolution that cannot exceed the
scale imposed by the Gaussian smoothing.

For a real application, the discrete and noisy nature of the data
should be taken into account, \ie\ some of the steps leading to
eq.~\ref{eq:ps} has to be modified. For example, to maintain the
orthogonality of the spherical Bessel functions on a finite spherical
volume one has to impose appropriate boundary conditions (see Fisher
\etal\ 1995 \& Zaroubi \etal 1995 for examples). However, the main
hurdle for this direct approach to power spectrum estimation is the
noise contribution; this issue is deferred to a future work.

\subsection{Estimation of $\beta$}
\label{variance}

\subsubsection{Estimator}
\label{estimator}

Eq.~\ref{eq:v2ptd2pt} is the most general relation derived in this
paper. However, in order to use it to estimate the value of $\beta$,
it is simpler to restrict ourselves to the relation between the
density and velocity variances, namely, apply the equation in the
limiting case of $\br= 0$ to yield:
\begin{equation}
{d\sigma^2_v(R_s) \over dR_s} = -{2 \over 3} \beta^2 { H_0^2}
R_s \sigma^2_\delta(R_s).
\label{eq:sigvsigd}
\end{equation}
Where $\sigma^2_v$ and $\sigma^2_\delta$ are the peculiar velocity and
density-contrast variances, respectively.

The numerical calculation of ${d\sigma^2_v(R_s) / dR_s}$ is
straightforward. One has to smooth the measured velocity field with a
Gaussian window, calculate its variance and obtain its derivative by
finite differencing (see subsection~\ref{noise} for a similar explicit
calculation). The right-hand-side of equation~\ref{eq:sigvsigd} is
obtained from the galaxy redshift catalogue by taking the variance of
the smoothed real-space density field.

The proposed estimator requires no heavy data manipulation and is easy
to calculate. Due to the smoothing involved, the estimator is robust
with regard to instabilities caused by the large random noise. In
addition, to avoid the cosmic variance contribution to the error
analysis, the comparison between the two types of data sets is
performed within the same region of space. Both features, simplicity
and stability, render the estimator very appealing to use.

\subsubsection{Noise}
\label{noise}

The contribution of the measurement error to the estimator in
equation~\ref{eq:sigvsigd} is readily calculated with the following
discrete approach. Let $\epsilon(\br_i)$ be the noise associated with
particle $i$, then the smoothed noise is,
\begin{equation}
\epsilon^S(\br_i)=\sum_l \epsilon(\br_l) W_{R_s}(\br_i-\br_l)
\label{eq:epssmooth}
\end{equation}

Subsequently, the expectation value of  the noise two point correlation is,
\begin{equation}
\langle \epsilon^S(\br_i) \epsilon^S(\br_j)\rangle 
=  \sum_l \langle\epsilon^2(\br_l)\rangle W_{R_s}(\br_i-\br_l) W_{R_s}(\br_j-\br_l). 
\end{equation}
The last equation assumes that the measured errors are statistically
uncorrelated.

We now require that $\br_i=\br_j$ and sum over all the data
points. The expectation value of the noise contribution to the
variance of the velocity is:
\begin{equation}
\sigma^2_N(R_s) = {1\over N}\sum_{i,l} \langle \epsilon^2(\br_l) \rangle
W^2_{R_s}(\br_i-\br_l).  
\label{eq:signoise}
\end{equation}
Therefore, the noise variance that adds to the left hand side of
equation~\ref{eq:sigvsigd} is readily obtained by finite differencing:

\begin{equation}
{d\sigma^2_N(R_s)\over dR_s} \approx {\sigma^2_N(R_s+\Delta R_s)
-\sigma^2_N(R_s)\over \Delta R_s}.
\end{equation}
The contribution of the noise variance to the right hand side of
eq.~\ref{eq:sigvsigd} is typically small and is neglected here.
However, it is straightforward to account for in the case of unusually
noisy data.


\section{Comparison between the SEcat and the 
PSC$\bmath{\lowercase{z}}$ catalogues}
\label{SEcatPSCz}

In this section equation~\ref{eq:sigvsigd} is employed for comparison
between the PSC$z$ galaxy redshift catalogue (Saunders \etal 2000,
Branchini \etal 2000) and the SEcat galaxy peculiar velocity catalogue
(Zaroubi \etal 2002) which is a combination of the two homogeneous
peculiar velocity catalogues, the SFI catalogue of spiral galaxies
(Giovanelli \etal 1998, Haynes \etal 1999) and the ENEAR catalogue of
early-type galaxies (da Costa \etal 2000). The SEcat catalogue extends
to a distance of about $70 \hmpc$ and the PSC$z$ goes to about twice
of that. Therefore, in order to avoid cosmic variance contamination of
the measurement the comparison between the two is restricted to the
closer distance.

Prior to applying the method to the actual data, however, one needs to
address the question of whether it is realistic to expect a reliable
estimation of the value of $\beta$ with noisy and close by catalogue
such as the SEcat. Hence, the next subsection is dedicated to
testing with mock catalogues how robust our estimator is.

\subsection{Testing with mock data}

The density and peculiar velocity mock catalogues used in this section
are derived from the $3.2\hmpc$ resolution reconstruction of the
density field from the PSC$z$ galaxy redshift catalogue (Branchini
\etal 2000), where the peculiar velocity field is obtained using
linear theory from the galaxy redshift space positions assuming a
value of $\beta=0.5$.  The mock SEcat peculiar velocity catalogue has
the same distances and number of points the real SEcat has, but with
the velocities of the PSC$z$ reconstructed velocity field. Obviously,
it would have been better to use a full nonlinear N-body simulation
with which to test the method. However, since the positions of the
actual measured velocities are controlled by the specific distribution
of the galaxies in the nearby universe we choose to test the method
with data that has the same spatial distribution as the real universe,
albeit the lack of full nonlinearity. Given the heavy smoothing
involved in the analysis this way of assigning velocities to mock data
is satisfactory -- for testing the method with full nonlinear
simulation see section~\ref{6dF}. After assigning the velocities to
the noise free mock data, we generate 30 mock catalogue with the
random errors added to their distance and velocity values in
concordance with the observational uncertainties.

\begin{figure}
\setlength{\unitlength}{1cm} \centering
\begin{picture}(8,9)
\put(-2.3, -3.){\includegraphics{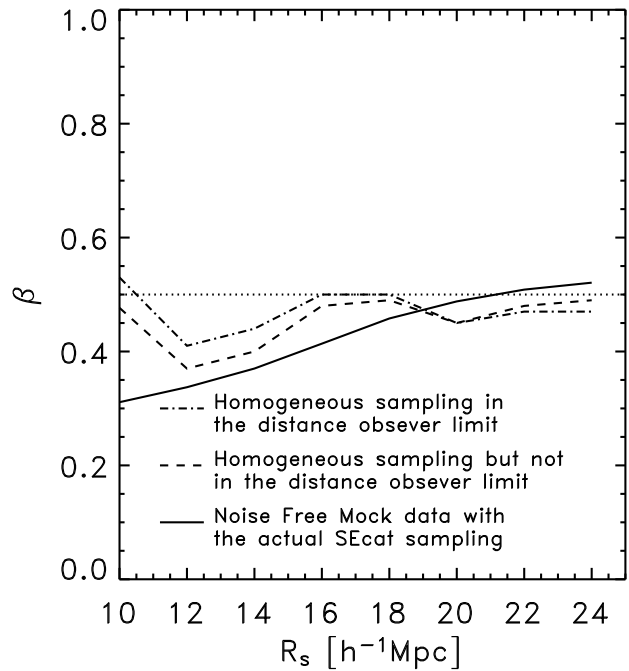}}
\end{picture}
\caption{$\beta$ as deduced from mock velocity data (as taken from
PSC$z$ high resolution data) designed to test various selection
effects. The dotted-dashed curve shows $\beta$ as calculated from a
homogeneously sampled velocity catalogue in the
distant-observer-limit. The dashed line shows beta from a
homogeneously sampled velocity catalogue but with the actual volume
coverage of SEcat, \ie\ the distant-observer-limit requirement is
relaxed. The solid line is $\beta$ as deduced from noise free mock
peculiar velocity data with the same selection effects as SEcat. The
dotted line shows the correct value of $\beta$}
\label{fig:mocktests}
\end{figure}

However, as a first step we wish to test whether the method works in
the distant observer limit with homogeneous sampling and noise free
data. In this case we have constructed the mock velocity data by
sampling the PSC$z$ catalogue on every 8-th grid point where the
velocity is taken to be equal to the $z$ component to mimic the
distance observer limit. Figure~\ref{fig:mocktests} shows $\beta$ as a
function of the smoothing scale (dotted-dashed line) which agrees
quite well, especially at larger smoothing scales, with the expected
value shown with the dotted horizontal line.

Next, the same mock data is used but the peculiar velocity is chosen
to be the radial velocity, \ie\ the distant-observer-limit is
relaxed. The points that were chosen for the comparison are restricted
to the range $30<r<60 \hmpc$ from the center of the box (relaxing this
restriction alters the results but marginally). The result of this
test is shown as a dashed line in figure~\ref{fig:mocktests}
indicating that the recovered $\beta$ is in agreement with its
original value.

The third issue to test is whether the spatial coverage of the data
set is sufficient, namely, whether the number of SEcat galaxies and
their actual sky distribution are good enough for a recovery of the
$\beta$ value? The answer is given by the solid line in
figure~\ref{fig:mocktests} clearly showing that at small smoothing
scales, $\beta$ is underestimated, then it increases with the
smoothing radius until the correct value is recovered on scales larger
than $18 \hmpc$. Here the positions of the mock velocity data are the
same as the galaxy positions in the SEcat catalogue but their
velocities are taken from the PSC$z$ velocity field with no noise
addition.

One might argue that there is no clear convergence of the value of
$\beta$ at $R_s = 24 \hmpc$ in figure~\ref{fig:mocktests}, therefore,
one might need to go to larger scales. However, given the size of the
current velocity data catalouges a larger scale smoothing becomes
comparable to the volume of the data set itself and one has to start
to worry about sampling issues within the Gaussian kernel itself. As
will be shown later, the lack of convergence is an issue, however less
severe, for the real SEcat and PSC$z$ data.

\begin{figure}
\setlength{\unitlength}{1cm} \centering
\begin{picture}(8,9)
\put(-2.3, -3.){\includegraphics{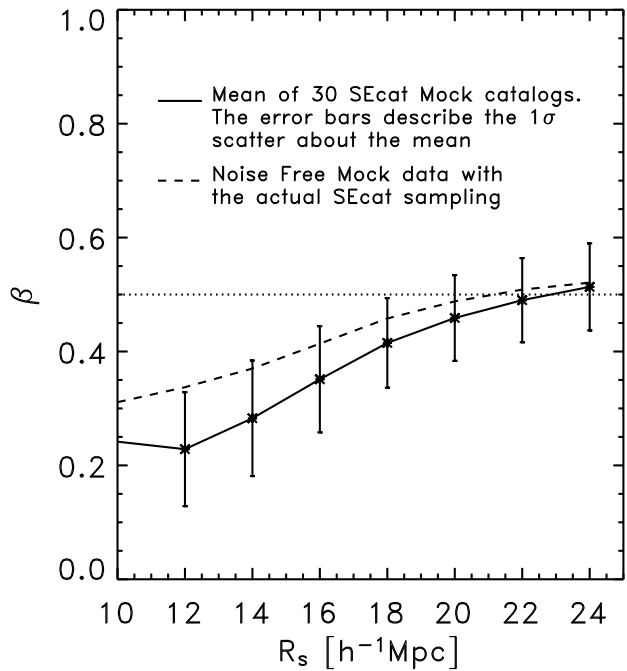}}
\end{picture}
\caption{$\beta$ as deduced from mock SEcat data where the underlying
$\beta$ in this catalogues is $0.5$ (dotted line). The solid line
shows the mean $\beta$ as obtained from 30 SEcat mock catalogues with
the error bars reflecting the 1-$\sigma$ uncertainty around the
mean. The dashed line shows the recovered value of $\beta$ for the
noise free case; clearly, there is an underestimation of $\beta$ from
the noisy data at small smoothing radii.}
\label{fig:mockvsPSCz}
\end{figure}

The final step in our testing is to apply the method to a ``full''
mock SEcat catalogue (with noise and actual sampling).  The solid line
in figure~\ref{fig:mockvsPSCz} shows the mean value of $\beta$ as a
function of the smoothing radius as recovered from 30 mock SEcat
catalogues with the error bars indicating the $1\sigma$ scatter about
the mean. $\beta$ is clearly well reconstructed with large smoothing
radii. There is also some bias in the mean value of $\beta$ at the
smaller smoothing radii with respect to the $\beta$ obtained from the
noise-free data (dashed line), which is probably due to overestimation
of the noise variance at smaller scales. This is not a big worry as on
small scales the PSC$z$ catalogue used to produce the velocity data
has limited nonlinear evolution due to its poor resolution ($3.2
\hmpc$) and its velocities are purely linear.

Please note that the error bars shown here are correlated. The
uncertainty estimates made for this figure, and for the rest of the
figures in the paper, are based on one of the error bars and not their
combination.

\subsection{$\beta$ from the real data}

\begin{figure}
\setlength{\unitlength}{1cm} \centering
\begin{picture}(8,9)
\put(-2.3, -3.){\includegraphics{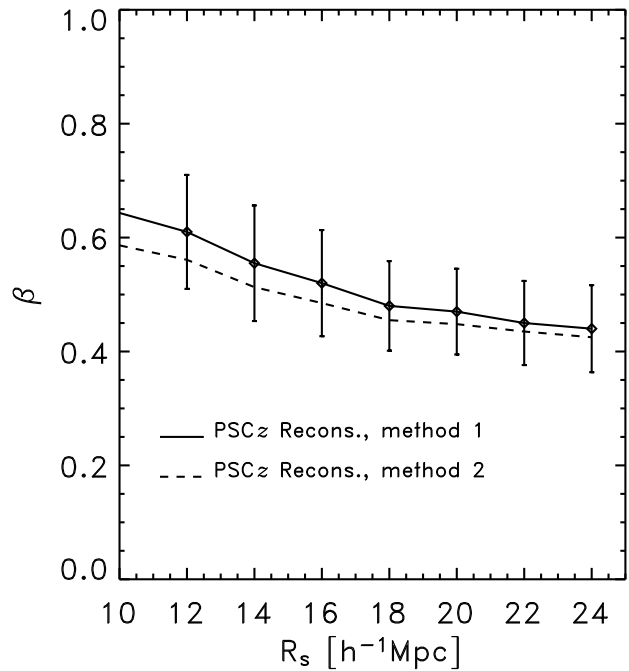}}
\end{picture}
\caption{$\beta$ as deduced from comparison of the real data SEcat
data with PSC$z$. The solid line and dashed line reflect results from
different assumptions from the density reconstruction. The error bars
in reflect the 1-$\sigma$ uncertainty.  }
\label{fig:SEcatvsPSCz}
\end{figure}

Having tested the method on mock catalogues and demonstrated that, on
large scales, it gives unbiased results for a SEcat-PSC$z$ comparison,
we now apply it to the real data. Figure~\ref{fig:SEcatvsPSCz} shows
the measured $\beta$ as a function of the smoothing radius where it
has a value of $0.6$ at smoothing radius of $10\hmpc$ but drops down
as the smoothing radius increases to $\approx 0.45$, the curve becomes
almost flat at $R_s \gsim 18\hmpc$. The error bars here are taken
from the $1\sigma$ uncertainties determined from the 30 mock
catalogues.

Branchini \etal\ (1999) have used two methods to solve for the
redshift distortion equation (Kaiser 1987) and reconstruct the
real-space density from the PSC$z$ galaxy redshift distribution, one
is based on the Yahil \etal\ (1989) iterative method and the other on
the Nusser \& Davis (1994) spherical harmonic expansion approach. In
the previous analysis we have used data obtained with the former
method. However, to examine the robustness of the measured value of
$\beta$ we perform the same comparison but with the later method. The
dashed line in figure~\ref{fig:SEcatvsPSCz} shows $\beta$ as a
function of smoothing radius deduced from the second method which is
well within the $1\sigma$ uncertainty level, albeit being slightly
smaller.

On small smoothing scales the behavior of the curves shown in
figure~\ref{fig:SEcatvsPSCz} is systematically different from those
obtained from the analysis of the mock catalogues, the former drops
with scale while the later increases.  We attribute this difference to
the fact that the PSC$z$ catalogue has a limited resolution and its
velocity field is purely linear.



\section{Future Surveys}

\label{6dF}

\subsection{Application to Mock 6dF catalogue}

In the near future, the 6dF Galaxy Survey (Jones \etal\ 2004) will
measure the redshifts of around 150000 galaxies, and the peculiar
velocities of a 15000-member sub-sample, over almost the entire
southern sky. When complete, it will be the largest redshift survey of
the nearby universe, reaching out to about $z \approx 0.15$, and more
than an order of magnitude larger than any peculiar velocity survey to
date.  Since the two datasets will be obtained from the same survey,
the galaxy redshift and peculiar velocity catalogues will have the
valuable attribute of being subjected to the same selection effects.

Despite the relatively large volume covered by the 6dF galaxy peculiar
velocity survey the relative nature of the errors in the $D_n-\sigma$
distance estimation might still diminish the information content of
the data. To evaluate this effect we apply eq.~\ref{eq:sigvsigd} to
mock 6dF galaxy redshift and peculiar velocity catalogues. In this
experiment the catalogues are constructed from the full nonlinear
N-body numerical simulation described by Cole \etal\ (1998),
specifically, the simulation labeled $\mathrm{L3S}$ in their
paper. The simulation assumes a CDM power spectrum of fluctuations
with $\Omega_m=0.3$, $\Lambda=0.7$, rms fluctuation of the mass
contained in spheres of radius $8\hmpc$, $\sigma_8=1.13$ and a CDM
power spectrum shape parameter, $\Gamma$, of $0.25$. The simulation
box side is $345.6\hmpc$ and has $192^3$ particles. The mock
catalogues where produces by carving out 6 hemispheres of radius
$150\hmpc$ of the simulation box. We obtain the redshift and peculiar
velocity catalogues with uniform sampling of the galaxies in the
simulated hemisphere in accordance with the expected sampling of the
6dF survey. The real-space distribution is presumed to have negligible
errors; but the distances in the peculiar velocity catalogues carry
errors of $20\%$ of their actual values. The input linear bias factor,
$b$ is one.

\begin{figure}
\setlength{\unitlength}{1cm} \centering
\begin{picture}(8,9)
\put(-2.1, -2.5){\includegraphics{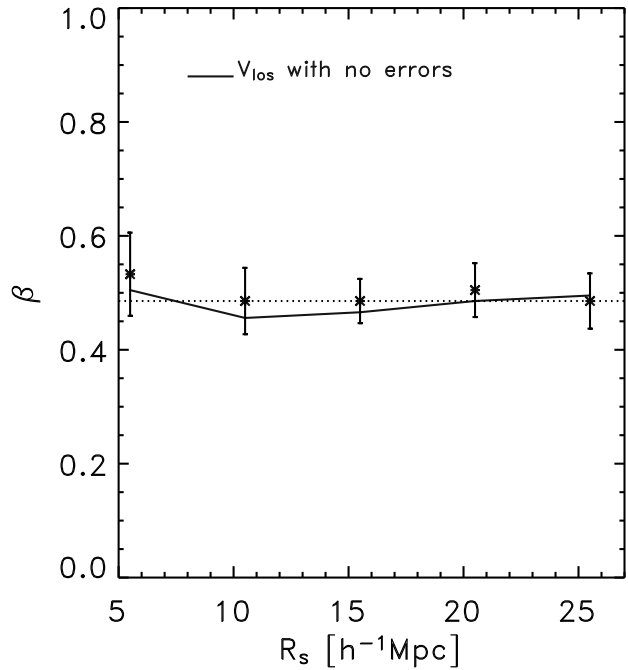}}
\end{picture}
\caption{$\beta$ as deduced from mock 6df catalogue as a function of
 smoothing radius. The stars show the mean results from 12
 independent data sets and the error bars show the scatter about it.
 The solid line shows the result obtained from the same analysis
 performed using a set of noise free-mock data. The dotted line is the
 input value of $\beta$.}
\label{fig:mock6df}
\end{figure}

The star symbols in figure~\ref{fig:mock6df} show the average $\beta$
value recovered from the 12 mock catalogs, as a function of smoothing
scale.  The error bars show the associated variance.  The continuous
line shows the average $\beta$ value recovered from the same 12 mock
catalogues in which no errors have been added to velocities.  Clearly,
the recovered $\beta$ is close to its input value at all smoothing
scales. If the error level we get is realistic the accuracy with which
the 6dF galaxy survey will recover the $\beta$ parameter ($\approx
0.05$) is indeed encouraging.

The recovery of $\beta$ down to $5\hmpc$ scale is very encouraging too
as it indicates that the Gaussian cell ``energy-like'' equation holds
also for the quasilinear regime. Obviously, this point needs to be
further explored with many simulations and over a wide range of point
separations.

\subsection{Kinematic and Thermal SZ Clusters}

Inverse Compton scattering of cosmic microwave background (CMB)
photons off thermal electrons within the hot intra-cluster medium of
galaxy clusters produce two effects. First, distortion of the CMB
black-body spectrum causing the cluster to appear brighter or dimmer
at different frequencies and, second, an achromatic modification of
its surface brightness. These effect are known, respectively, as the
thermal and kinematic Sunyaev-Zeldovich effects (Sunyaev \& Zeldovich
1972). The two combined with a measure of the cluster temperature give
the cluster's radial peculiar velocity component to a high degree of
accuracy. Current estimates of the measured distance-independent {\it
absolute} uncertainty are as low as  $130 \kms$ (Holder 2004).

The thermal component of the SZ effect is now routinely measured with
interferometers and major efforts are underway to survey the sky with
in the thermal SZ relevant spectral range. Since the SZ effect is
redshift independent, this kind of survey will provide an unbiased
catalogue of the massive clusters as far back as their formation redshift
(see \eg\ Carlstrom, Holder \& Reese 2002 for a review).

The kinematic SZ is an order of magnitude weaker than the thermal
component and therefore has been harder to measure (Holzapfel \etal\
1997; Benson \etal\ 2003). In the future however, the kinematic SZ
effect will be measurable and together with the thermal SZ component will
provide wide angle surveys of galaxy-cluster peculiar velocities
up to redshift of about $2$. Such data will probe the evolution of
the dark energy and galaxy-cluster bias evolution and
clearly distinguish between various theoretical scenarios of
cosmological evolution.

\begin{figure*}
\setlength{\unitlength}{1cm} \centering
\begin{picture}(8,8.8)
\put(-3, -1.6){\includegraphics{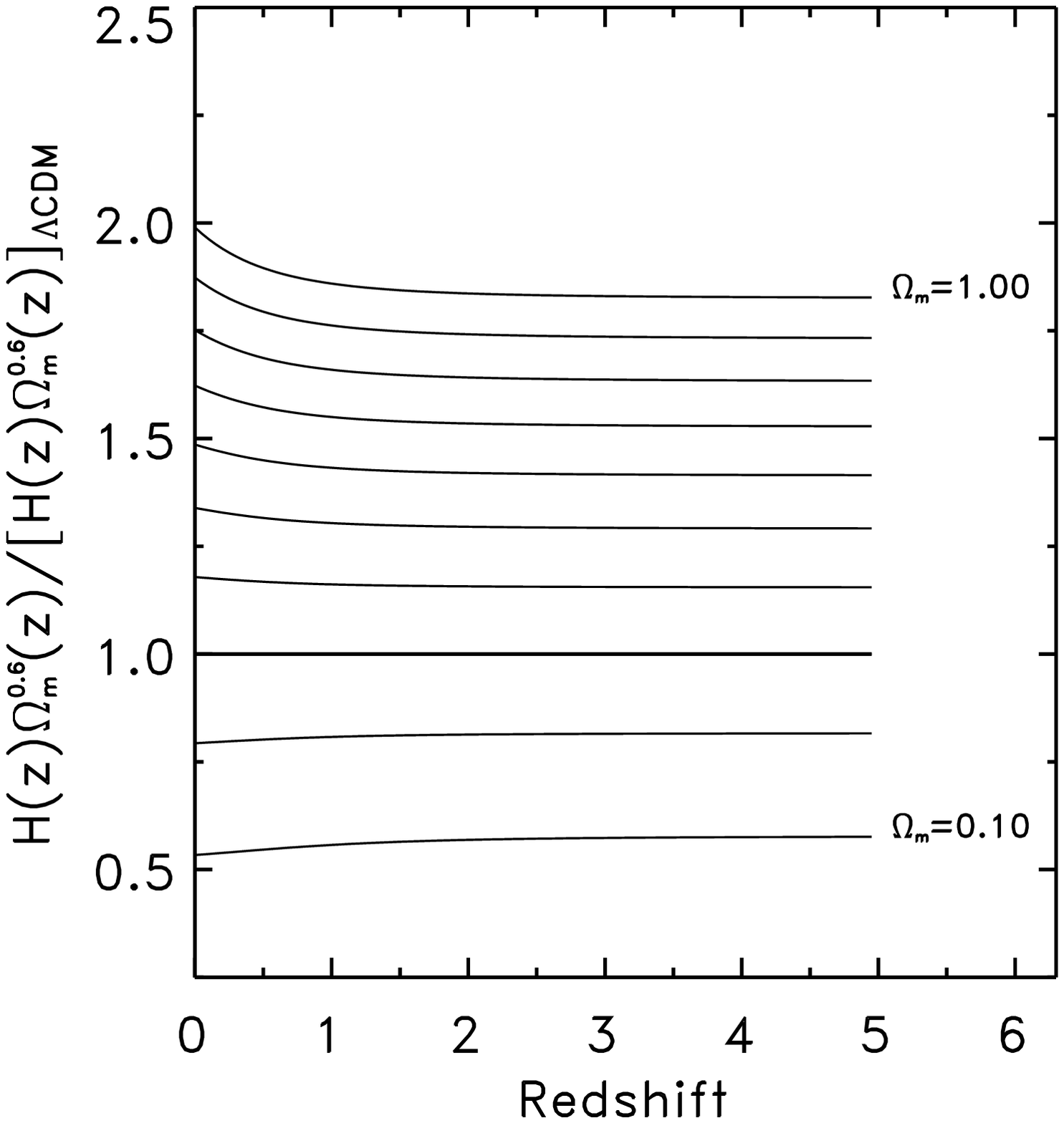}}
\end{picture}
\begin{picture}(8,8.8)
\put(-1.5, -1.6){\includegraphics{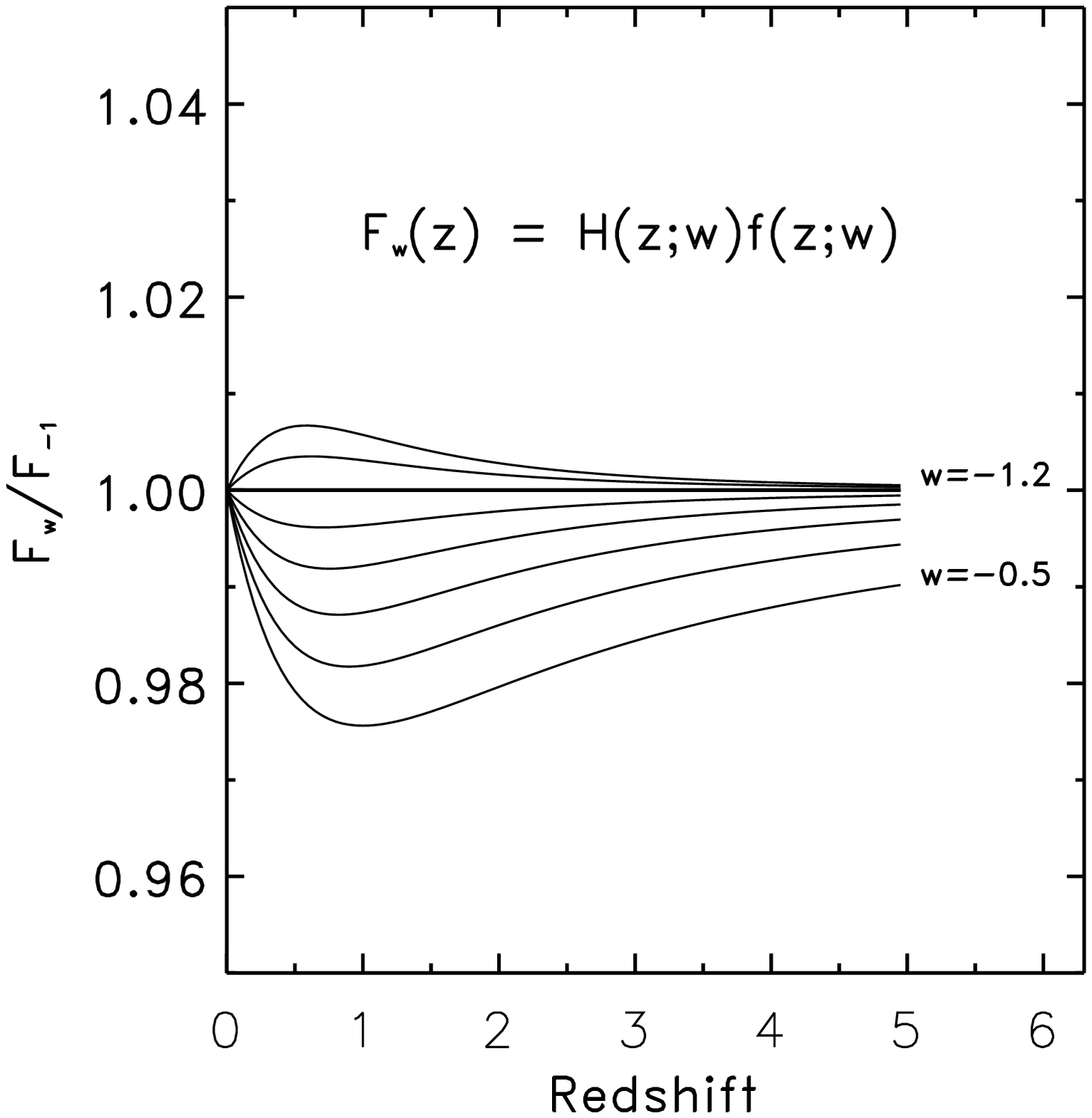}}
\end{picture}
\caption{The left hand panel shows the evolution of $H f(\Omega_m)$ as
 a function of redshift for flat cosmological $\Lambda$CDM models with
 $\Omega_m=0.1,0.1,\dots,1$, relative to the $\Omega_m=0.3$ and
 $\Omega_\Lambda=0.7$ case. The right hand panel shows the evolution of
 the same quantity as a function of redshift for different dark energy
 equations of state assuming a flat $\Lambda$CDM universe with
 $\Omega_m=0.3$ and $\Omega_\Lambda=0.7$. The lines are for the
 dark-energy equation of state parameter, $w$, values of $\{-0.6,
 -0.7, -0.8, -0.9, -1. -1.1, -1.2\}$, with the thick line showing the
 evolution in the $w=-1$ case.  Although, the magnitude and extremum
 location of the ratio vary with $w$, the F function itself depends
 very weekly on the value of $w$.}
\label{fig:dark}
\end{figure*}

Like the 6dF galaxy survey, the future SZ surveys will probe the
density and the peculiar velocity of the same region of space with the
same objects and therefore allow a measurement of $\Omega_m$ (through
$\beta$).  The left hand panel of figure~\ref{fig:dark} shows the
evoution of $H f(\Omega_m)$ as a function of redshift for different
values of $\Omega_m$ for flat $\Lambda$CDM universes normalize to the
case of $\Omega_m=0.3$ and $\Omega_\Lambda=0.7$ case. This figure
demonstrates the sensitvity of the cluster SZ peculiar motions to the
value of $\Omega_m$.

The right hand panel of figure~\ref{fig:dark} shows the evolution of
$H f(\Omega_m)$ as a function of redshift for various values of the
dark energy equation of state parameter, $w$, in a flat universe
(Haiman, Mohr \& Holder 2001). As pointed out by Lahav \etal (1991)
the evolution of $f(\Omega_m)$ partially cancels out with the
evolution in the Hubble parameter. The weak dependence of the
evolution on $w$ clearly shows that the equation of state is very hard
to measure with peculiar velocity data. On the other hand however,
this insensitivity facilitates a very accurate measurement of the
clusters biasing factor and its evolution as a function of redshift.

The expected superior quality of the measured peculiar velocity of
individual clusters is hampered by their sparseness. Therefore, it is
essential to analyze the data with methods that are stable with
respect to this feature. The method developed here is a good candidate
as it is simple, easy to apply, involves no complicated inversion
schemes and the vast majority of the measured clusters will satisfy
the distant-observer-limit assumed in the derivation. Initial
application of the method to realistic mock catalogues shows a good
success in the recovery of the $\beta$. However, the mock data to
which we applied it were at redshift zero and limited in size. When
the cluster mass cut-off exceeds $8\times10^{13} M_{\odot}$ the
simulation box is left with very small number of clusters. In order to
test the applicability of the method properly one should apply it to
very large scale simulations that span the redshift range of $0-2$; a
task that will be deferred to the future.


\section{discussion}
\label{discussion}

In this paper we introduced the {\it Gaussian cell two point
``energy-like'' equation} connecting the two point density and
peculiar velocity correlation functions. The interpretation of this
equation is that the change in the velocity correlation function is
caused by density variation coming from scales larger than the scale
set by the Gaussian smoothing; this analytic cancellation of the small
scale power is particular to Gaussian kernels. Two practical
applications of the {\it Gaussian cell two-point energy-like equation}
have been developed here, the first is direct matter power spectrum
estimator from peculiar velocity data, and the second is $\beta$
measurement from comparison of galaxy peculiar velocity and redshift
surveys. The later application was restricted to the velocity
dispersion, \ie\ the $r=0$, case.

In the $r=0$ case, the relation derived here is similar in its
mathematical form to the Irvine-Layzer cosmic energy equation. This is
not surprising as each of the two relations reflect some sort of
energy balance and should, due to dimensionality arguments, be homologous.

Restricting the main formula to the variance case the relation could
be easily used to estimate the value of $\beta$ from comparison
between galaxy peculiar velocity and redshift catalogues. In this
paper we showed that despite their proximity the PSC$z$ galaxy
redshift survey and the SEcat galaxy Peculiar velocity data could be
reliably used to derive the value of $\beta$. The result is consistent
with that of previous analyses. The variance case has also been shown
to apply to the 6dF galaxy survey, despite being far from the
distant-observer-limit. Using mock 6dF catalogues we have demonstrated
that our method can be successfully used to extract cosmological
parameters from the real sample.

In the future, the eminent detectability of the kinematic
Sunyaev-Zeldovich effect will provide peculiar velocity measurements
for large number of galaxy clusters at redshifts extending back to the
formation epoch of cluster ($z \approx 2$). This type of data is ideal
to explore with the {\it Gaussian cell two-point energy-like equation}
as it satisfy all of the required assumptions and have small
measurement errors with large spatial coverage. The redshift coverage
of the SZ data will allow an accurate measurement of the evolution of
the clusters biasing factor with redshift. The main hurdle these data
sets will pose is the limited resolution with which they will sample
the universe as the comoving rms distance between rich galaxy-clusters
is of the order of $30\hmpc$.

We have also shown that the {\it Gaussian cell two-point energy-like
equation} could be used to estimate the matter power spectrum peculiar
velocity data in a non-parametric fashion. This is a very important
application since the current measurements of the mass power spectrum
from peculiar velocity employs likelihood analysis with specific
models that almost certainly do not properly account for the noise
contribution (Zaroubi \etal 1997, 2001). A non-parametric measurement
on the other hand will allow a scale-by-scale dissection of the
various components contributing to the measured power spectrum
allowing the isolation of the noise part.

Finally, given the simplicity of equation~\ref{eq:v2ptd2pt} it is
tantalising to attempt to extend it to the quasi-linear regime to
obtain a nonlinear description of the evolution of the two point
peculiar velocity correlation function similar to the very successful
quasi-linear extension of its density counterpart (Hamilton \etal\
1991). Indeed, figure~\ref{fig:mock6df} gives an encouraging
indication that the equation might hold for the quasi-linear
regime. This will be further explored in a future work.

\section*{acknowledgments}
The authors would like to acknowledge the hospitality of the
``Institute of Theoretical Physics'', the Technion, Haifa. We also
would like to thank Adi Nusser for his very insightful comments.

{}

\end{document}